\documentclass[12pt,preprint]{aastex}

\newcommand{\dechms}[4]{$#1^{\rm h}#2^{\rm m}#3\mbox{$^{\rm s}\mskip-7.6mu.\,$}#4$} 
\newcommand{\decdms}[4]{$#1^{\circ}#2'#3\mbox{$''\mskip-7.6mu.\,$}#4$} 
\newcommand{\mdeg}[2]{$#1\mbox{$^\circ \mskip-7.6mu.\,$}#2$} 
\newcommand{\msec}[2]{$#1\mbox{$''\mskip-7.6mu.\,$}#2$}

\shorttitle{Cyg OB2 \# 5: resolving the nature of the non-thermal source.}
\shortauthors{Ortiz-Le\'on et al.}

\begin{document}

\title{The non-thermal, time-variable radio emission from Cyg OB2 \# 5: A wind-collision region}

\author{Gisela N. Ortiz-Le\'on\altaffilmark{1}, 
Laurent Loinard\altaffilmark{1}, Luis F. Rodr{\'{\i}}guez\altaffilmark{1}, 
Amy J. Mioduszewski\altaffilmark{2}  \\ and Sergio A. Dzib\altaffilmark{1}}

\email{g.ortiz@crya.unam.mx}


\altaffiltext{1}{Centro de Radiostronom\'ia y Astrof\'isica, Universidad Nacional Aut\'onoma de Mexico,
Morelia 58089, Mexico}
\altaffiltext{2}{National Radio Astronomy Observatory, 1003 Lopezville Road, Socorro, NM 87801,
USA}

\begin{abstract}
The radio emission from the well-studied massive stellar system Cyg OB2 \#5 is known to
fluctuate with a period of 6.7 
years between a low-flux state when the emission is entirely of free-free origin, and a
high-flux state when an additional non-thermal component (of hitherto unknown nature) 
appears. In this paper, we demonstrate that the radio flux of that non-thermal component 
is steady on timescales of hours, and that its morphology is arc-like. 
This shows that the non-thermal emission 
results from the collision between the strong wind driven by the known contact binary in the
system, and that of an unseen companion on a somewhat eccentric orbit with a 6.7-yr
period and a 5 to 10 mas semi-major axis. Together with the previously reported wind-collision 
region located about \msec{0}{8} to the north-east of the contact binary, Cyg OB2 \#5 appears
to be the only multiple system 
known so far to harbor two radio-imaged wind-collision regions.
\end{abstract}

\keywords{stars: individual (Cyg OB2 \#5) --- stars: winds --- radio continuum: stars}

\section{Introduction}

Cyg OB2 \#5 (V729 Cyg, BD +40 4220) is a radio-bright early-type multiple system comprising 
an eclipsing contact binary with a 6.6-day orbital period.
In addition to the primary radio 
component coincident with the contact binary, Abbott et al. (1981) and 
Miralles et al. (1994) reported on 
the existence of a radio ``companion'' located $0\rlap.{''}8$ to the north-east of the contact binary. 
Contreras et al. (1997) determined that this secondary radio component has an elongated shape and 
lies in-between the binary system and a B-type star, first reported by Herbig (1967). 
The emission from the secondary component is non-thermal (i. e. synchrotron;
White 1985) and has been ascribed to the interaction region formed 
by the wind from the contact binary and that of the B-type star.
These interaction regions consist of a contact discontinuity with two shocks, one 
shock on either side of the contact discontinuity.
The study of objects such as Cyg OB2 \#5 
allows to investigate the physics of particle acceleration in massive stars environments
(De Becker 2007; Benaglia 2010). The study of sources of cosmic ray acceleration
(in addition to supernova remnants) is very important since there is observational evidence that
they are produced in a variety of Galactic sources (Adriani et al. 2011). 

Observations of Cyg OB2 \#5 taken over a period of about 20 years show that the radio 
emission varies in time between a low- and a high-flux state (Persi et al. 1990;
Contreras et al. 1997). In the low-flux 
state, the emission has a flux density of about 2 mJy at 4.8 GHz, and is due to thermal free-free 
emission from the ionized material in the stellar wind driven by the contact binary. The spectral
index of the emission is $0.60 \pm 0.04$ (Kennedy et al. 2010), as expected for a wind with a density 
decreasing as $1/r^2$ (Panagia \& Felli 1975; Olnon 1975;
Wright \& Barlow 1975). During the high-flux state, the total flux rises to about 8 mJy at 4.8 
GHz, as an additional non-thermal component appears. Because of the non-thermal nature
of the additional component, the spectral index of the system is flatter in the high-flux state 
than during the low-flux state.

Recently, Kennedy et al. (2010) reanalysed all existing VLA observations of Cyg OB2 \#5 and showed 
that the radio flux variations arise directly in the primary component with a periodicity of 
$6.7 \pm 0.2$ 
yr, while the flux from the secondary component to the north-east remains constant in time. These 
authors proposed that the variations can be represented by a simple model in which a source (with 
constant non-thermal emission) moves around the contact binary in a $6.7$-yr orbit  and the varying 
radio emission results from the variable free-free opacity of the wind between the source and the 
observer (Kennedy et al. 2010).  The non-thermal emission requires the presence of a fourth star in the 
system, but two different scenarios are possible. In the first one, the non-thermal emission would
arise from a wind-collision region between the contact binary and the fourth star in the system. 
The fourth star would evidently have to be sufficiently massive to drive a wind powerful enough 
to produce an interaction region with the wind from the contact binary.
The presence of a wind-collision region is supported by the 
detection of X-rays (Waldron et al. 1999; Linder et al. 2009). Interestingly,  Kennedy et al. (2010) 
found evidence in the radial velocities of the contact binary for a reflex motion resulting from a 
putative star with a mass of  $23^{+22}_{-14}\, \mbox{M}_\odot$, consistent with a late O or early 
B type star. Alternatively, the fourth star itself could generate the non-thermal emission as a 
consequence 
of coronal activity. This would favor a scenario where the fourth star in the system would be a 
low-mass 
object since convection in the stellar interior is required for coronal activity to occur.

Two diagnostics can be used to distinguish between these two scenarios. The first one is
related to variability. Coronal radio emission 
from low (Feigelson \& Montmerle 1999) and some intermediate-mass 
(Dzib et al. 2010) stars tends to be very variable, and to exhibit short-duration flares 
on timescale of 
hours (Osten \& Wolk 2009). If the non-thermal source in Cyg OB2 \#5 is from a star 
with coronal activity, 
variability on timescales of hours would be expected to occur during the high-flux state, when the 
star travels in front of binary system. In the low-flux state, the free-free opacity of the wind 
from the contact binary would hide the orbiting non-thermal source and we would solely  detect the 
constant, thermal emission from the wind. In the wind interaction scenario, on the other hand, no
rapid variations are expected in the high-flux state since wind collision regions are relatively
extended and generate fairly
constant radio emission, provided timescales significantly shorter than the orbital
period are considered. The second diagnostic is related to the size of the non-thermal emitting 
region. For Cyg OB2 \#5 at 1.7 kpc (Torres-Dodgen et al. 1991), the 
emission due to a wind interaction is expected 
to have an extent of order 10 mas (Pittard \& Dougherty 2006), whereas emission from the active 
magnetosphere of a low-mass star would be more compact that  0.1 mas. Rodr{\'{\i}}guez et al. (2010)
estimated 
an angular size of $\leq 0\rlap.{''}02$ for the non-thermal source by subtracting the persistent 
thermal component (low-flux state) from the high-flux state (thermal plus non-thermal emission 
present). Both scenarios remain consistent with this estimate, so higher angular resolutions are needed.

In this paper, we first re-analyze (in Section \ref{sec:1}) a total of 19 archival observations obtained 
during the high-flux state of Cyg OB2 \#5 between 1983 and 2003, to search for short timescale
flux variability. Detection of such variability would support the scenario where the non-thermal emission
comes from a low-mass companion, whereas a non-detection would be in better agreement with the
wind interaction scenario. In the following section (\ref{sec:2}), we present the 
first Very Long Baseline Interferometry
(VLBI) observation ever reported of Cyg OB2 \#5. This observation will enable us to put direct 
constrains on the morphology and size of the non-thermal emission in Cyg OB2 \#5.

\section{A Search for Short-Timescale Radio Variability}\label{sec:1}

The data analyzed here comprise 19 VLA observations of Cyg OB2 \#5 at frequencies between 1.4 and 
43.3 GHz during the high-flux state of the source, and were obtained from the NRAO archive. The 
on-source integration time for these observing sessions varies between 11.5 to 157.7 minutes. The 
detailed parameters of the observations are given in Ortiz-Le\'on (2010). The archive data were edited 
and calibrated using the software package Astronomical Image Processing System (AIPS) of NRAO. The 
absolute amplitude calibrator for 16 of the 19 epochs was 1331$+$305, while 0137$+$331 was used 
for the remaining three epochs. The phase calibrator was 2007$+$404 for all epochs. The adopted and 
bootstrapped flux densities of the amplitude and phase calibrators are given in Ortiz-Le\'on (2010).

These observations have been analyzed by several authors (e.g.\ Persi et al. 1990; Kennedy et al. 2010)
but using the entire on-source integration time of each session as a single data point. Here,
we are interested, instead, in the short timescale variability. Our strategy to search for such
variability was the following. First, images of Cyg OB2 \#5 were made with the AIPS task IMAGR 
for each observation epoch. With the AIPS task JMFIT, the precise position of the primary component 
was determined. The $(u,v)$ data was then recentered at the position of the main component using 
the AIPS task UVFIX. With these new $(u,v)$ data, the real and imaginary parts of the interferometer data were 
plotted as a function of time, averaged over the $(u,v)$ plane. The real part gives us
information on the flux density of the source and the imaginary part on its position and
symmetry. We then averaged over time in bins from 
5 to 20 minutes. A detailed description of this technique is given by Neria et al.\ (2010).
To avoid contamination from extended emission, in observations at 4.8 GHz with the D 
configuration, we used only $(u,v)$ data with baselines larger than 0.125 km
(suppressing the emission from structures larger than $\sim2'$). Furthermore, a self-calibration 
process was applied in four epochs to diminish remaining phase errors in the visibility.

\begin{figure}[!t]
  \centerline{\includegraphics[height=0.5\textwidth,angle=0]{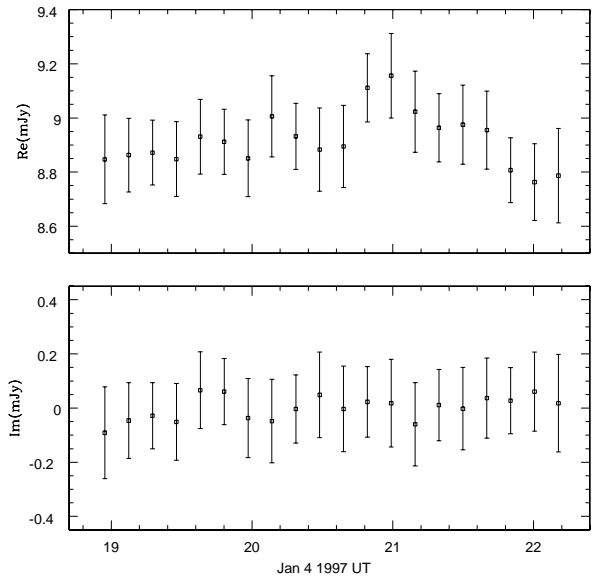}}
 \caption{Real (top) and imaginary (bottom)
 components of the interferometer data at 4.8 GHz for 1997
 January 4 as a function of universal time (UT). 
 The real component is consistent with no significant variability
 over the period of the observation and a constant flux density of
 $8.9 \pm 0.1$ mJy.
 The imaginary component is consistent
 with zero, indicating that the source is symmetric about
 the phase center (the origin of the visibility plane) and
 has no significant structure on these spatial scales.
 \label{fig_visibilities}}
 \end{figure}

We searched for rapid variability in the light curves of the real part of the visibility, but found 
no statistical evidence for variations at typical levels of 10\%. An example of this data
analysis is shown in Figure 1. This lack of strong short-timescale 
variability clearly favors the scenario where the non-thermal radio emission comes from a wind
interaction region. It should be pointed out, however, that while fast variability is common in 
non-thermal stellar sources (i. e. Feigelson \& Montmerle 1999; G\"udel 2002; Bower et al.\ 2003; 
Forbrich et al.\ 2008; G\'omez et al.\ 2008), there is at least one example known of a young
stellar source generating steady non-thermal emission (Andr\'e et al.\ 1988, 1991; Loinard et al. 
2008). Thus, the lack of fast variability cannot be considered as a definitive proof that
the non-thermal emission comes from a wind interaction region. A direct, high-angular resolution 
image is needed to distinguish between a compact stellar source and a wind-collision region.
We will present such an image in the coming Section.

\begin{figure}[!t]
  \centerline{\includegraphics[height=0.5\textwidth,angle=-90]{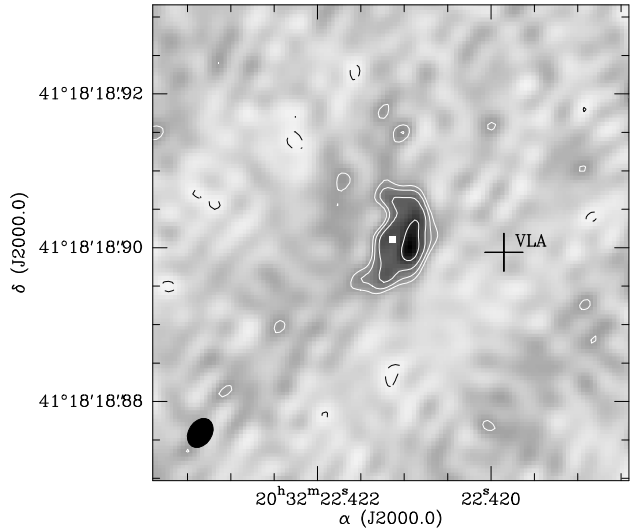}}
  \caption{Image of Cyg OB2 \#5 obtained with the VLBA on December 13, 2010.
  The contours are at --1, 1, $\sqrt{2}$, 2, 2$\sqrt{2}$ times the 3$\sigma$ noise
  level in the image (0.165 mJy beam$^{-1}$). The synthesized beam is shown
  at the bottom left. The position of the contact binary at the epoch of this observation
  is shown as a cross, whose size represents the error bar. The small white square is an
  estimate of the position of the additional star in the region. \label{fig_vlba}}
\end{figure}

\section{VLBA Observations of Cyg OB2 \#5}
\label{sec:2}

Cyg OB2 \#5 was observed at $\lambda$ = 3.6 cm ($\nu$ = 8.4 GHz) with the Very Long 
Baseline Array (VLBA) on December 13, 2010.  The position of the peak of the 
source was at $\alpha_{J2000.0}$ =
\dechms{20}{32}{22}{421}, $\delta_{J2000.0}$ = +\decdms{41}{18}{18}{90}.
The observation was 5 hours, and consisted of 
series of cycles with two minutes spent on source, and one minute spent on the 
main phase-referencing quasar J0218+3851, located \mdeg{3}{6} away.  Every
15 minutes, the radio bright X-ray binary Cyg X-3, which is $20'$ from Cyg OB2 \#5 
was also observed.  Although Cyg X-3 is very close to Cyg OB2 \#5 it was not used 
as a primary calibrator due to its extreme variability in both flux density and 
morphology and the fact it is heavily scattered at the observing frequency 
(Desai \& Fey 2001).
The data were edited and calibrated using the AIPS software 
following standard procedure for phase-referenced VLBA observations. The 
calibration determined from the observations of J0218+3851 was applied to both 
Cyg OB2 \#5 and Cyg X-3.  At this point most of the phase errors left are caused 
by the \mdeg{3}{6} separation between the source and the phase calibrator.  To 
remove most of this, Cyg X-3 was self-calibrated and the phase gains
determined from that self-calibration were then interpolated and applied to Cyg 
OB2 \#5.  This resulted in a significant improvement in the quality of the image 
of Cyg OB2 \#5 compared to a direct phase transfer between J0218+3851 and Cyg OB2 
\#5.  Cyg OB2 \#5 is quite resolved so the best image was made by limiting 
the maximum uv length to 60000 k$\lambda$ and using pure natural weighting.

The resulting image is shown in Figure 2. The total flux density of the source is 
$\sim$5 mJy, in agreement
with the value expected for the non-thermal component as derived
from lower angular resolution observations. This result suggests that
at these high angular resolutions we are observing synchrotron emission.
The peak flux density of the source is 1.1 
mJy beam$^{-1}$, implying a brightness 
temperature, $T_B \geq 1.6 \times 10^6$ K. This brightness temperature
is not large enough to clearly favor a non-thermal origin and VLBI observations
at several wavelengths are needed to establish the nature of the emission beyond
doubt.
The source is clearly resolved with a total
extent, mostly along the north-south direction, of about 10 mas (consistent with the upper
limit reported by Rodr\'{\i}guez et al.\ 2010). The arc-like morphology of the emission is 
the structure expected for a wind interaction region. Since the apex of the arc shape is
located to the west, one would expect the source with the strongest wind (the contact
binary system in the present case) to be located west of the arc. To examine this 
issue, the position of the contact binary at the epoch of the VLBA observations must
be determined.

\begin{figure}[!t]
  \centerline{\includegraphics[height=0.5\textwidth,angle=0]{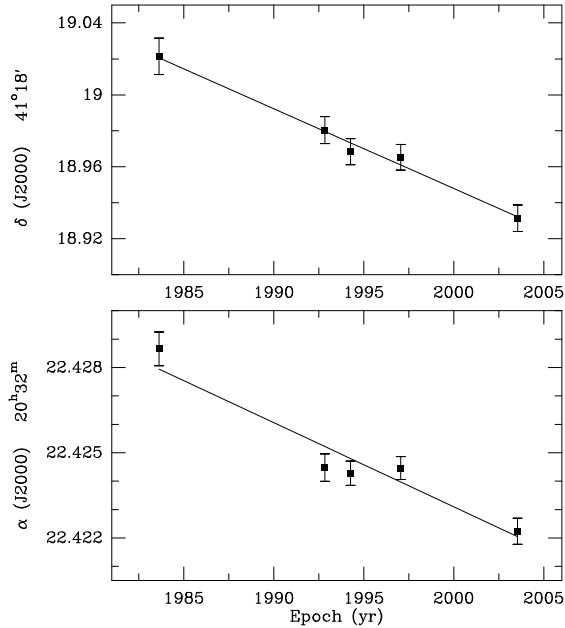}}
  \caption{Position in right ascension (top) and declination (bottom) of the contact 
  binary in Cyg OB2 \#5 measured with the VLA at five epochs between 1983 and
  2004. The best fit to these data with a uniform proper motion is shown as solid
  lines.  \label{fig_vla}}
\end{figure}

From the VLA observations mentioned in Section \ref{sec:1}, we selected the 3.6 and 
6 cm data obtained in the A configuration of the array. Using these observations, the
position\footnote{All observations used the same phase calibrator (2007+404) and were
registered to the most recent coordinates of that source. The relative astrometry of these
data and the VLBA observations is better than a few mas.} of the radio emission associated 
with the contact binary was determined at 5 epochs between 1983 and 2004 (Figure 3)
and its proper motion was measured to be:

\begin{eqnarray}
\mu_\alpha\cos \delta  & = &  -3.80 \pm 0.71 \mbox{~mas~yr$^{-1}$}\nonumber\\
\mu_\delta  & = &  -4.43 \pm 0.32 \mbox{~mas~yr$^{-1}$.} \nonumber
\end{eqnarray}

\noindent
This is in good agreement with the proper motions of several Wolf Rayet stars
located in the same region (Dzib \& Rodr\'{\i}guez 2009). The position of the source
at the epoch (2010.95) of the VLBA observation and its error bar is shown as a cross
on Figure 2. Within the errors, this position lies along the direction of the apex of the
arc seen in the VLBA image, as expected. The star responsible for the second colliding 
wind must be located east of the arc along its bisector, and presumably fairly close to 
the arc itself since its wind is likely to be significantly less powerful than that of the contact 
binary. Assuming that the wind momentum rate of the star responsible for the second colliding
wind is 0.1 that of the contact binary and following Cant\'o et al. (1996),
we obtain a location at the position of the small white square shown in Figure 2.

The detection of an arc-like structure in the VLBA observations demonstrates that the
non-thermal emission detected during the high-flux state of Cyg OB2 \#5 most probably
originates in a wind collision region, but raises 
an interesting additional issue. The separation
between the wind collision region and the position of the contact binary (Figure 2) is
only about 12 mas. The diameter of the region surrounding the contact binary where
the free-free radiation is emitted was measured to be about 
60 mas (Rodr\'\i guez et al. 2010). Thus, the
wind-collision arc appears to be located within the nominal region occupied by the 
free-free emission. The free-free opacity in that region should prevent the non-thermal
emission from escaping and reaching the observer. A possible solution to this apparent 
conundrum is that the wind driven by the contact binary is very inhomogeneous
(Blomme et al. 2010; Muijres et al. 2011), and consequently
partially transparent at centimeter wavelengths. This is a problem that has to be studied
in the future because the non-thermal excess emission is clearly seen even
at frequencies as low as 1.4 GHz (Persi et 
al. 1985; Ortiz-Le\'on et al. 2011), where the free-free opacity is
expected to be quite large.

Assuming that our estimate of the position of the fourth star in the system is reasonable, 
the current separation between that star and the contact binary is about 15 mas, or 
25.5 AU (at a distance of 1.7 kpc; Torres-Dodgen et al.\ 1991). If this provided a 
good estimate of the orbital semi-major axis, then
using Kepler's third law the total mass of the system would
have to be about 400 M$_\odot$. This seems unreasonably large, and suggests that 
the orbit is somewhat eccentric, and the system currently near apastron. For instance,
taking an eccentricity of 0.5 could reduce the mass of the system by up to a factor 8, 
down to about 50 M$_\odot$, and consistent with a system of 3 O-type stars. 
A high eccentricity is in agreement with that deduced by Kennedy et al.
(2010) from their best fit model for the variable non-thermal flux from Cyg
OB2 \#5.

Interestingly, if for these estimates we adopt the shorter distance of 925 pc proposed
by Linder et al. (2009), we obtain a total mass of the system of only
about 60 M$_\odot$, assuming that the current separation between the
contact binary and the putative position of its companion is  
representative of
the orbital semi-major axis. We are currrently monitoring the emission of Cyg OB2 \#5 
with VLBI observations and may be able to provide the geometric parallax and an
accurate distance in the future.

\section{Conclusions}

In this paper, we have first re-analyzed archival VLA observations of the radio emission
from Cyg OB2 \#5 taken when the system was in its high-flux state. This re-analysis
shows that the flux of the non-thermal emission in the system does not suffer
strong short-term variability. We then presented the first VLBI image of this system,
and demonstrated that the non-thermal component most probably originates in an arc-shaped 
wind-collision region resulting from the interaction between the wind driven by the
known contact binary in the system, and that powered by a companion star in a 
6.7-yr, somewhat eccentric orbit. VLBI observations at several
wavelenghts are needed to establish unambiguously the nature of the emission
from the arc-shaped region. To our knowledge, Cyg OB2 \#5 is the
only multiple system 
known so far to harbor two radio-imaged wind-collision regions,
one described in this paper,
and the other reported previously by Contreras et al.\ (1997).

\acknowledgments
We thank our refere, Micha\"el De Becker, for his valuable comments that significantly improved
this paper.
G.O.L, L.L., L.F.R., and S.D\  acknowledge the financial support of DGAPA, UNAM and CONACyT, 
M\'exico. L.L.\ is grateful to the Guggenheim Foundation for financial support. The National Radio 
Astronomy Observatory is a facility of the National Science Foundation operated under cooperative 
agreement by Associated Universities, Inc.

\clearpage

\end{document}